\documentclass[journal=jacsat,manuscript=article]{achemso}

\usepackage[version=3]{mhchem}
\usepackage{amsmath,amssymb,mathrsfs,accents}
\usepackage{graphicx}
\usepackage{siunitx}

\title{Attosecond-Stable Two-Dimensional Spectroscopy by a Sagnac-Based Modulating System and a sub-4-fs Continuum Source}

\author{Wei-Chung Feng}
\altaffiliation{These authors contributed equally to this work.}
\affiliation{Institute of Photonics Technologies, National Tsing Hua University, Hsinchu 300044, Taiwan}

\author{Bo-Han Chen}
\altaffiliation{These authors contributed equally to this work.}
\email{bohan.chen@ee.nthu.edu.tw}
\affiliation{Institute of Photonics Technologies, National Tsing Hua University, Hsinchu 300044, Taiwan}

\author{Chih-Hsuan Lu}
\affiliation{Institute of Photonics Technologies, National Tsing Hua University, Hsinchu 300044, Taiwan}

\author{Howe-Siang Tan}
\affiliation{School of Chemistry, Chemical Engineering and Biotechnology, Nanyang Technological University, Singapore 637371, Singapore}

\author{Shang-Da Yang}
\email{shangda@ee.nthu.edu.tw}
\affiliation{Institute of Photonics Technologies, National Tsing Hua University, Hsinchu 300044, Taiwan}

\author{Kai Chen}
\email{kai.chen@vuw.ac.nz}
\affiliation{Robinson Research Institute, Victoria University of Wellington, Lower Hutt, Wellington 5010, New Zealand}
\alsoaffiliation{The Dodd-Walls Centre for Photonic and Quantum Technologies, Dunedin 9016, New Zealand}
\alsoaffiliation{MacDiarmid Institute for Advanced Materials and Nanotechnology, Wellington 6012, New Zealand}

\abbreviations{2DES,MPC,TA}

\begin{document}

\begin{abstract}
We present a two-dimensional electronic spectroscopy (2DES) platform driven by a novel Coherent Loop-based Integrated Modulating and Beamsplitting System (CLIMBS). Coupled with an octave-spanning multiple-plate continuum (MPC) source, CLIMBS enables broadband, phase-coherent measurements with attosecond-level time delay precision. Its Sagnac-inspired, nearly common-path geometry provides exceptional long-term phase stability without active feedback, eliminating beam walk-off and preserving beam pointing during delay scans. Delay calibration using spectrally resolved interferometric fringes yielded a wedge angle in excellent agreement with the designed geometry, confirming precise, linear coherence time control. The MPC technique generates broadband excitation pulses spanning 550--980 nm and temporally compressed to 3.7 fs. This bright, few-cycle source enables simultaneous interrogation of widely separated electronic and vibronic transitions, with high temporal and spectral resolution, allowing 2DES to capture vibronic cross peaks, energy-transfer pathways, and undistorted ground-state bleaching (GB), stimulated emission (SE), and excited-state absorption (ESA) features across a broad spectral window. System performance was benchmarked on chlorophyll-\textit{a} in methanol, where the excitation bandwidth fully covers the $Q_x$ and $Q_y$ bands, ensuring distortion-free spectra. The nearly collinear configuration of CLIMBS eliminates beam walk-off during delay scanning, supports ultrabroadband few-cycle 2DES enabled by the high-brightness MPC source, and maintains attosecond-level phase stability, providing a simple and robust platform for high-fidelity multidimensional spectroscopy.
\end{abstract}

\section{Introduction}

Coherent two-dimensional electronic spectroscopy (2DES) is a powerful method for interrogating photoinitiated processes, including energy transport, exciton delocalization, charge-transfer formation, system-bath interaction, linewidth broadening, and vibronic coupling \cite{Tekavec:09}. 2DES spreads congested optical responses into a second frequency dimension, enabling excitation-frequency-resolved detection. One of the most common 2DES implementations is the pump--probe beam geometry \cite{Tekavec:09}. Compared with conventional transient absorption spectroscopy (TAS), pump--probe 2DES uses a pair of phase-locked pulses with inter-pulse delay $\tau$ as the pump. Upon Fourier transform over $\tau$, 2DES achieves Fourier-limited temporal resolution while simultaneously maintaining high spectral resolution along both the excitation and detection axes \cite{10.1063/1.1398579}. This multidimensional capability reveals excitation--emission correlations, isolates distinct nonlinear pathways, and uncovers coherent dynamics inaccessible to pump--probe techniques.

Despite its powerful capabilities, 2DES faces two principal experimental challenges: producing broadband few-cycle pulses with adequate spectral coverage and maintaining attosecond-level phase locking between a pair of pump pulses to avoid spectral distortions along the excitation axis. Among the established approaches for generating broadband ultrafast pulses, noncollinear optical parametric amplification (NOPA) offers broad tunability and sub-10-fs durations \cite{Steves:19}, but its output efficiency is relatively low, phase-matching conditions constrain its usable bandwidth, and its noncollinear geometry can introduce spatiotemporal distortions such as spatial chirp and angular dispersion \cite{Zaukevicius:11}. Gas-filled hollow-core fibers (HCF) provide higher conversion efficiency and excellent spatial mode quality \cite{Ma:16}. However, they typically require a large optical footprint, are highly sensitive to beam pointing and in-coupling efficiency, and depend on gas handling and pressure control, making them less compact and more difficult to operate in routine 2DES experiments. These trade-offs motivate the pursuit of broadband pulse-generation schemes that are compact, high-brightness, and stable. In this context, the multiple-plate continuum (MPC) technique provides distinct advantages that make it particularly well suited for two-dimensional electronic spectroscopy. MPC can generate an intense, octave-spanning supercontinuum with high pulse energy (tens of $\mu$J) and excellent shot-to-shot stability, allowing both pump and probe beams to be derived from a single broadband source. MPC-based sources have been readily applied to transient absorption spectroscopy to enhance bandwidth and temporal resolution \cite{Tamming2021_ScientificReports_12847}, demonstrating their robustness and high beam quality. These characteristics make the MPC architecture an attractive platform for broadband, few-cycle excitation in 2DES experiments where simultaneous temporal resolution and spectral coverage are essential.

Having addressed the challenge of generating broadband few-cycle excitation, the second key requirement for high-fidelity 2DES is maintaining attosecond-level phase stability between the pump pulses. The excitation-frequency axis of a 2D spectrum is obtained by Fourier transforming the signal along the coherence time, meaning that even slight phase fluctuations distort the coherence and lead to broadened peaks, reduced cross-peak contrast, or shifts in the excitation axis. This requirement becomes increasingly difficult to satisfy at higher optical frequencies because the shorter optical period makes the pulses far more sensitive to phase fluctuations \cite{doi:10.1021/ar900227m}. Conventional interferometric schemes, such as the Michelson interferometer, generate pulse replicas by varying the optical path length in one arm, but precise delay tuning demands ultra-high-precision mechanical stages; for instance, a 1 fs delay corresponds to a displacement of only about 150 nm for a round trip. Moreover, because the two pulses propagate along separate arms, Michelson geometries are highly susceptible to environmental perturbations such as vibration, air currents, and thermal drift, often necessitating active stabilization. These limitations make traditional two-arm interferometers poorly suited for delivering the long-term phase stability required for broadband 2DES, and their use in 2D spectroscopy is limited to the infrared \cite{shim2025frontier}.

To improve delay precision, several alternative approaches have been developed. The Translating-Wedge-Based Identical Pulses eNcoding System (TWINS) is an elegant configuration that introduces optical delay via two orthogonally oriented birefringent wedge pairs \cite{Brida:12}. Due to its common-path geometry, TWINS offers exceptional stability and sub-femtosecond delay resolution, and has been successfully implemented in two-dimensional spectroscopy across multiple spectral ranges by selecting appropriate birefringent materials \cite{Rehault:14,Borrego-Varillas:16}. However, the use of birefringent materials may introduce significant dispersion, potentially stretching or distorting the pulse replicas. Moreover, TWINS is subject to several practical limitations, including additional dispersion from the birefringent wedges, a restricted range of accessible coherence times $\tau$, beam walk-off arising from birefringent wedges, and the inability to independently control the polarization of the two pulse replicas.

Another technique involves rotating optical flats to exploit refractive-index variations for fine delay adjustment, achieving sub-femtosecond precision \cite{doi:10.1021/acs.jpca.0c00285}. Nevertheless, this method offers a limited scanning range and demands synchronized rotation of identical flats to prevent beam displacement. Alternatively, a pulse shaper, or acousto-optic programmable dispersive filter (AOPDF), can also be employed to generate pulse replicas \cite{Myers:08}, providing inherent stability due to its common-path configuration. However, implementing a pulse shaper increases system complexity and may impose limitations on the maximum scannable coherence time, spectral bandwidth, and transmission efficiency.

In prior 2DES implementations, Sagnac interferometers were used for interferometric signal detection rather than pump-pulse generation \cite{Courtney:14}. Here, we develop a novel Sagnac-based pulse shaper, termed the Coherent Loop-based Integrated Modulating and Beamsplitting System (CLIMBS), which produces two phase-locked collinear replicas without relying on birefringent or polarization optics. Within the CLIMBS setup, the two counter-propagating pulses are spatially separated, allowing each pulse to be manipulated independently, such as in polarization control. At the same time, their passage through identical optical elements ensures attosecond-level phase stability. The double-pass geometry further preserves spatial overlap and output pointing during delay scans. Due to its inherently collinear output configuration, CLIMBS can be easily integrated into a conventional transient absorption setup, enabling straightforward implementation of phase-locked pump--probe measurements. This design combines the stability of a nearly common-path interferometer with the flexibility of independently tunable pulse characteristics, while supporting a long coherence-time range, eliminating beam drift during delay scanning, and supporting ultrabroad spectral bandwidths required for few-cycle excitation. Together, these features provide a robust platform for broadband phase-coherent multidimensional spectroscopy.

\section{Experimental Section}

\begin{figure}[ht]
    \centering
    \includegraphics[width=0.7\linewidth]{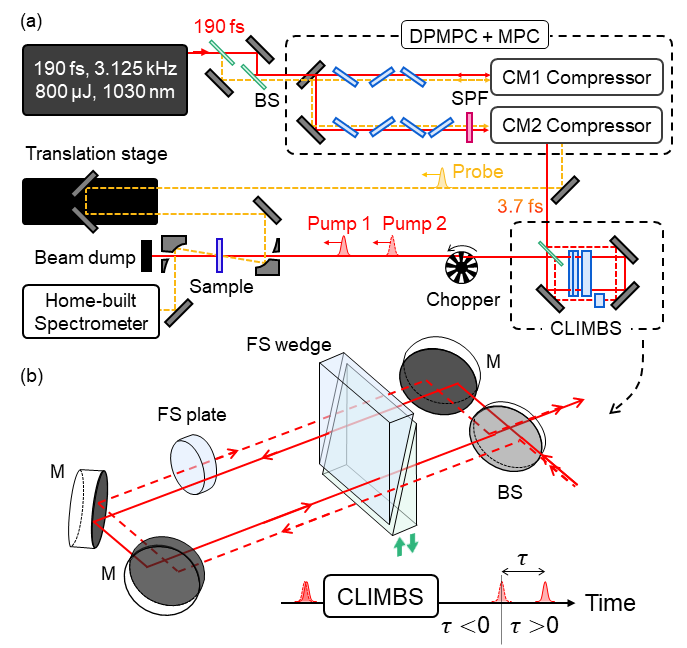}
    \caption{
    (a) Schematic setup of the 2DES platform featuring the multiple-plate continuum (MPC) source and CLIMBS. BS: beam splitter; MPC: multiple-plate continuum; DPMPC: double-pass multiple-plate continuum; SPF: 980 nm short-pass filter; CM Compressor: chirped-mirror compressor. (b) Detailed arrangement of the the CLIMBS module. The input beam is split into counter-propagating components (solid and dashed lines) that are spatially separated and circulate the inner and outer loops. The temporal delay $\tau$ between the pulses in the two loops is controlled via the insertion of a wedge pair. The two beams are finally recombined and exit the pulse shaper collinearly. This common-path geometry ensures attosecond-level phase stability and makes the system easily integrable with pump--probe or transient absorption configurations. BS: beam splitter; M: mirrors; FS: fused silica. The inset illustrates the resulting pair of time-delayed pump pulses produced at the CLIMBS output.
    }
    \label{fig:setup}
\end{figure}

Our experimental setup is shown in Figure~\ref{fig:setup}(a). The setup of our multiple-plate continuum (MPC) light source follows our previous work \cite{Tamming2021_ScientificReports_12847}. Briefly, a commercial Yb:KGW amplifier (Pharos, Light Conversion) provided 190 fs, 3.125 kHz, 2.5 W (\SI{800}{\micro\joule} per shot) pulses at 1030 nm. The output was split by a low-GDD 50/50 beam splitter to generate two identical pulses that were sent through a double-passed multiple-plate compression (DPMPC) \cite{chen2022double} and a single-pass MPC \cite{Lu:19} for nonlinear spectral broadening. A short-pass filter with a 980 nm cut-off was inserted to remove the fundamental part at 1030 nm and its red spectral component. The dispersion before the sample was then compensated by eight bounces off chirped mirrors (Ultrafast Innovation). The compressed pulses at the sample position exhibited a 3.7 fs full width at half-maximum (FWHM) duration, verified by polarization-gating frequency-resolved optical gating (PG-FROG), as shown in Figure~S1. These two compressed pulses serve as pump and probe pulses in the downstream transient absorption and 2DES measurements.

The pump pulse was routed into a CLIMBS module to generate a collinear pump pulse pair. A detailed arrangement of the CLIMBS module is illustrated in Figure~\ref{fig:setup}(b). At the entrance of the pulse shaper, a low-GDD broadband 50:50 beam splitter (UFBS5050) was employed to split the pump pulse and launch counter-propagating replicas into inner and outer loops. A thin pair of fused-silica wedges with an apex angle of $3^\circ$ was inserted into the inner loop to precisely control the optical delay difference between the two loops. One of the wedges was mounted on a motorized translation stage (LS-2430, NATORS) with sub-nanometer positioning resolution, enabling excellent and repeatable delay control. Owing to the closed-loop geometry of the nearly common-path architecture, the inner beam traverses the wedge pair in both directions. This configuration inherently self-compensates for any lateral beam deviation, guaranteeing that the pulses emerging from the inner and outer loops remain strictly collinear and free of spatial walk-off. In addition to the wedge pair, the design also supports delay control using a single optical window on a rotational mount. In the outer loop, a 1.6 mm fused-silica plate provides a fixed optical delay, allowing for delay scanning across $\tau = 0$. To enable independent control of the two pump pulses, the input beam is intentionally shifted by a small amount so that the clockwise and counterclockwise beams become spatially separated while still sharing a nearly common optical path and passing through identical optical elements. At the output, the two beams are recombined to be collinear and fully overlapped, yielding a stable and well-controlled pump pulse pair.

After emerging from the CLIMBS module, the two pump pulses pass through an optical chopper synchronized with the laser and operating at half its repetition rate. Meanwhile, the probe pulse is directed to a linear translation stage (DL325, Newport) that provides a delay range of approximately 2.2 ns. Both the pump and probe beams are then focused onto the sample at a slight off-angle. Finally, a home-built spectrometer, synchronized with the laser, records the normalized differential probe transmission spectra ($\Delta T/T$) on a shot-by-shot basis for transient absorption or 2DES measurements.

\section{Results and Discussion}

\subsection{Calibration and Performance of the CLIMBS module}

\begin{figure}[htbp]
    \centering
    \includegraphics[width=0.7\linewidth]{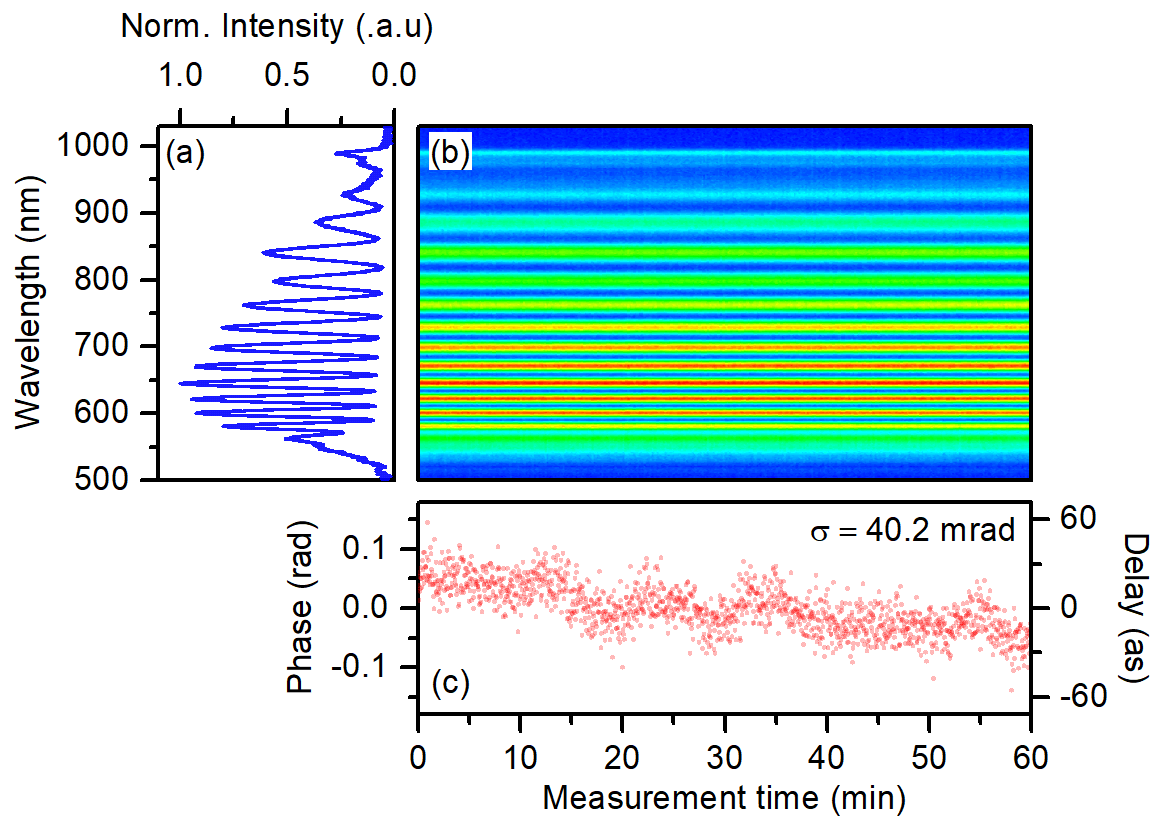}
    \caption{
        (a) High-contrast spectral interferogram obtained at a fixed delay $\tau$. 
        (b) Sequence of interferograms recorded continuously over 60 min, showing stable fringe phase with a standard deviation of $\sigma \approx 40~\mathrm{mrad}$ (16 as at 750 nm). 
        (c) Phase extracted from the measured interferogram at a fixed position, confirming the absence of phase drift or fringe fading. These results demonstrate the attosecond-level phase stability of the CLIMBS module.
    }
    \label{fig:stability}
\end{figure}

The stability of the CLIMBS was evaluated by recording the spectral interference between the two counter-propagating MPC pulses using a fiber-coupled commercial spectrometer (HR4000, Ocean Optics). Figure~\ref{fig:stability}(a) shows the spectral interference obtained at a fixed delay $\tau$. The high fringe visibility confirms that the pump pulse pair remain strict phase coherence after traversing the CLIMBS module. The persistence of well-defined fringes across the visible--NIR range further highlights the broadband capability of the MPC source and CLIMBS. Figure~\ref{fig:stability}(b) presents a series of spectral interference data acquired continuously over 60 min. The fringe phase remained stable within a standard deviation of $\sigma \approx 40~\mathrm{mrad}$, corresponding to a temporal jitter of approximately 16 as at a carrier wavelength of 750 nm, which confirms the absence of phase drift or fringe fading, as shown in Figure~\ref{fig:stability}(c). These results verify that CLIMBS provides attosecond-level delay stability and long-term reproducibility essential for interferometric multidimensional spectroscopy.

\begin{figure}[ht]
    \centering
    \includegraphics[width=0.7\linewidth]{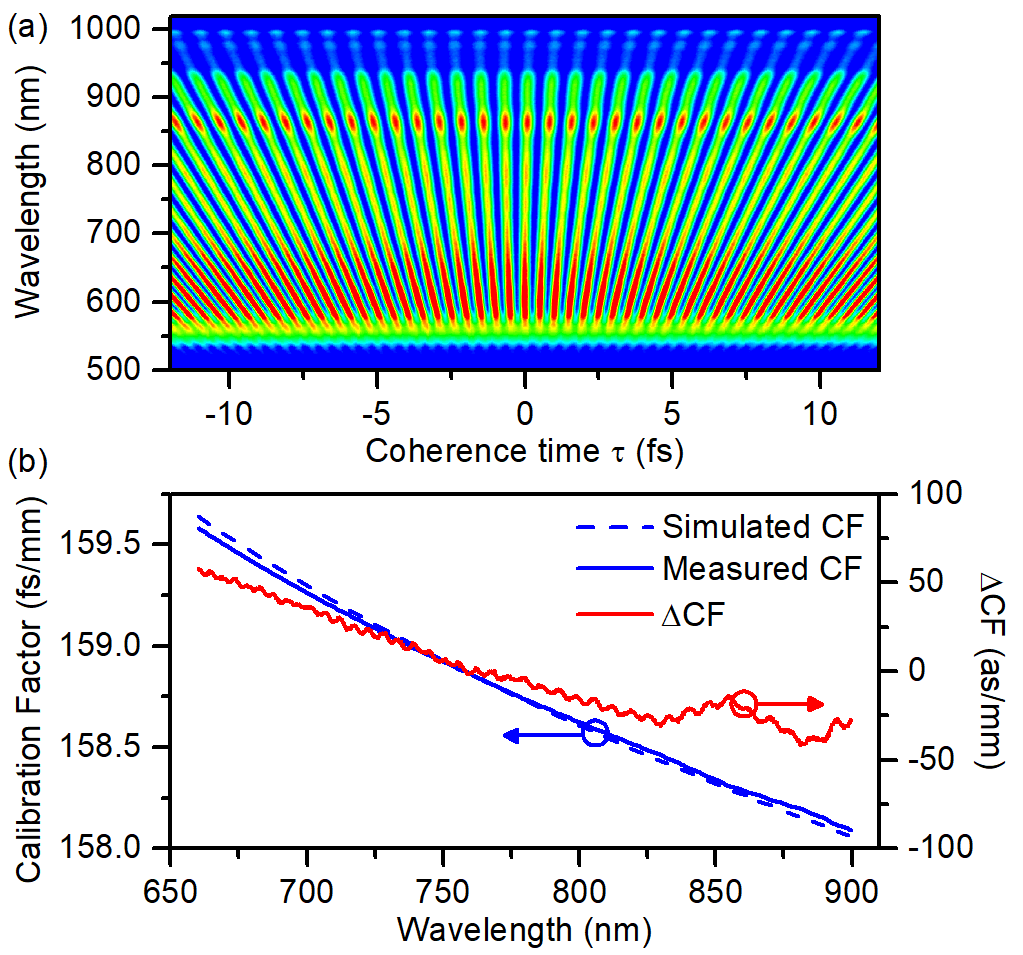}
    \caption{
       (a) Spectrally resolved interference fringes recorded as a function of coherence time $\tau$. The clear and periodic fringes indicate that CLIMBS provides smooth and highly precise temporal delay control over a broad spectral range. This configuration enables wide and repeatable scans with attosecond-level timing precision, essential for interferometric multidimensional spectroscopy. (b) Comparison between measured (solid blue) and simulated (dashed blue) calibration factors of the wedge-pair delay line as a function of wavelength. The red curve (right axis) represents the error $\Delta$CF between measured and simulated values. The excellent agreement verifies the accuracy of the calibration and confirms the high performance of the CLIMBS design.
    }
    \label{fig:scan}
\end{figure}

After verifying the long-term phase stability, the scanning performance of the pulse shaper was evaluated by translating one of the fused-silica wedges into the beam path of the inner loop of the CLIMBS. Because the beam traverses the wedge pair twice, once in each direction, the resulting optical path difference is effectively doubled. In this configuration, the coherence-time delay is generated by the refractive-index difference between air and fused silica, enabling a large and continuous delay range. In our setup, a maximum coherence-time delay of approximately 4.8 ps is achieved at a central wavelength of 700 nm. The wedge translation was computer-controlled to ensure smooth and repeatable motion across the entire travel range. Figure~\ref{fig:scan}(a) shows the spectrally resolved interference fringes recorded as a function of coherence-time delay $\tau$, obtained by applying an inverse Fourier transform to the interferograms. The distinct and periodic fringes demonstrate the high precision and linearity of the delay control across a wide spectral range. By analyzing the fringe periodicity as a function of wedge insertion, a wavelength-dependent delay calibration factor (CF) (fs$\cdot$mm$^{-1}$) was derived \cite{Zhu:19}, effectively mapping wedge translation to temporal delay. The experimentally obtained calibration factor [blue solid, Figure~\ref{fig:scan}(b)] was in good agreement with values simulated based on the dispersion of fused silica [blue dashed, Figure~\ref{fig:scan}(b)], yielding wedge apex angles of $3.0062^\circ$ and $3^\circ$, respectively. This excellent agreement confirms both the accuracy of the calibration and the high performance of the proposed CLIMBS.

\subsection{Demonstration of Broadband, Phase-Coherent 2DES Enabled by the MPC--CLIMBS Platform}

To validate the broadband response, phase stability, and temporal precision of the proposed MPC--CLIMBS platform, 2DES measurements were performed using chlorophyll~\textit{a} in methanol as a benchmark system. Chlorophyll~\textit{a} is ideally suited for this purpose because its well-separated $Q_x$ and $Q_y$ absorption bands and thoroughly studied ultrafast dynamics provide a stringent test of spectral coverage and phase fidelity \cite{KHYASUDEEN2019110480}.

\begin{figure}[htbp]
    \centering
    \includegraphics[width=0.7\linewidth]{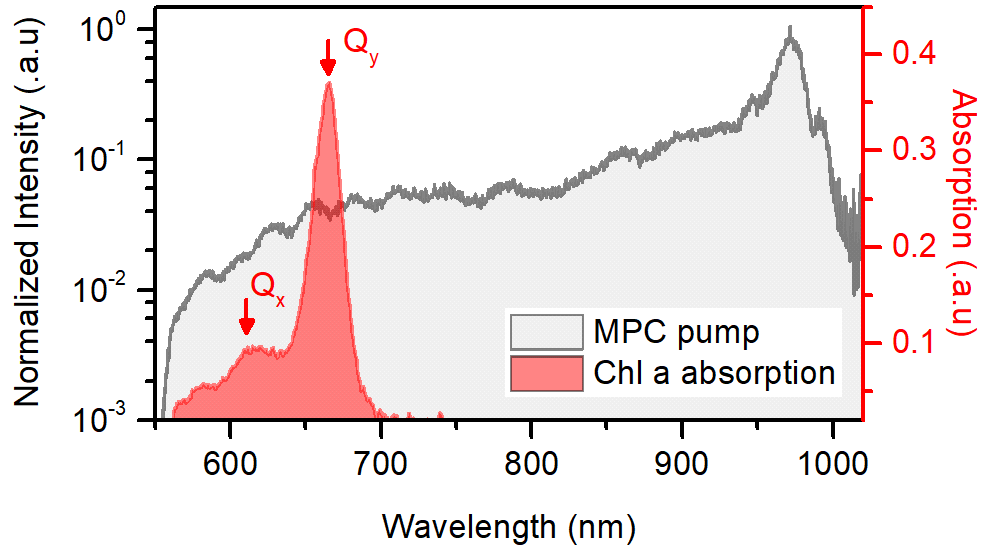}
    \caption{
    Normalized intensity spectrum of the MPC-generated pump pulse (gray shaded area, plotted on a logarithmic scale) overlaid with the absorption spectrum of chlorophyll~\textit{a} (red shaded area). The MPC spectrum fully overlaps with the $Q_x$ and $Q_y$ absorption bands, providing spectrally matched excitation for ultrafast transient absorption and two-dimensional spectroscopy measurements.
    }
    \label{fig:spectrum}
\end{figure}

Figure~\ref{fig:spectrum} illustrates the spectral overlap between the steady-state absorption of chlorophyll a and the MPC excitation spectrum generated by CLIMBS. The smooth, octave-spanning continuum (550--980 nm) fully covers both the $Q_x$ and $Q_y$ bands, providing sufficient spectral bandwidth to excite the entire absorption manifold without gaps. Such complete coverage is essential for faithful reconstruction of electronic couplings within a single measurement, as partial or uneven excitation can distort peak shapes and obscure the underlying frequency correlations \cite{Son:17}.

The coherent pump pair produced by CLIMBS has attosecond-level delay precision and long-term reproducibility, which are essential for maintaining coherence during the Fourier transformation along the coherence-time axis. The pulse characterization discussed in Section 2 shows that the excitation pulses are nearly transform-limited. This guarantees that the 2DES spectra reflect the intrinsic electronic coherence of the sample, rather than artifacts from pulse distortion or timing jitter.

\begin{figure}[htbp]
    \centering
    \includegraphics[width=0.7\linewidth]{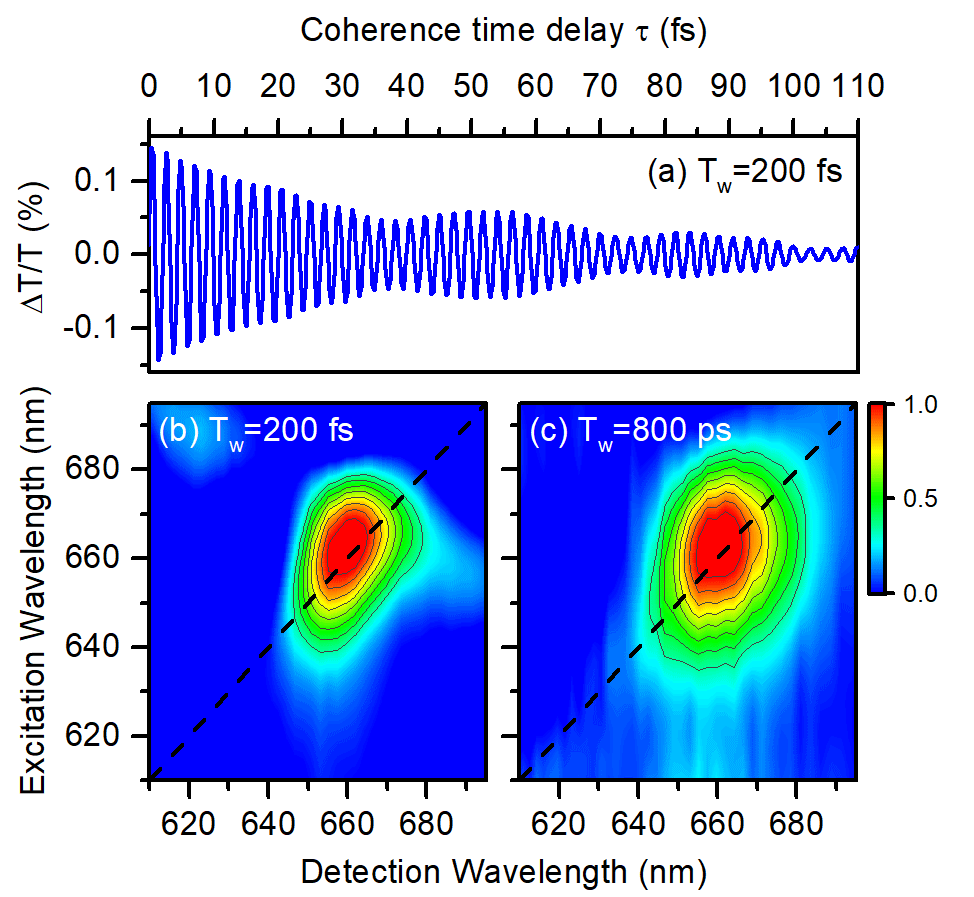}
    \caption{
    (a) Oscillating transient signal probed at 670 nm. 2D electronic spectrum of chlorophyll~\textit{a} recorded at (b) $T_\mathrm{w} = 200$ fs and (c) $T_\mathrm{w} = 800$ ps.
    }
    \label{fig:2des}
\end{figure}

Two 2DES spectra were recorded under magic-angle conditions ($54.7^\circ$ between pump and probe) \cite{schott2014generalized} at population times of 200 fs and 800 ps to capture both the early-time coherent response and the long-time population relaxation of chlorophyll~\textit{a}. As shown in Figures~\ref{fig:2des}(b,c), the 2D spectrum at the early population time of $T = 200$ fs exhibits a diagonally elongated shape. This elongation signifies a strong correlation between the excitation and detection frequencies during early population time. As the population time increases to $T = 800$ ps, the spectral shape evolves into a more symmetrical round feature. This morphological change arises from spectral diffusion, in which the detection frequency gradually loses its correlation with the initial excitation frequency. It originates from fluctuations in the chromophore transition frequency during the population time. This evolution from a diagonally elongated feature at early times to a more isotropic peak at long waiting time is consistent with previously reported 2DES behavior of chlorophyll~\textit{a} in solution \cite{KHYASUDEEN2019110480}. Although cross peaks are not observed in our measurements, the observed spectral-shape evolution confirms that the setup captures the expected dynamics while maintaining stability from the femtosecond to sub-nanosecond regime.

\section{Conclusion}

In summary, we have developed a broadband, phase-coherent two-dimensional electronic spectroscopy platform that integrates an octave-spanning multiple-plate continuum source with our novel CLIMBS architecture. The MPC light source provides bright, smoothly varying excitation spanning 550--980 nm and can be compressed to 3.7 fs, offering both the broad spectral coverage and few-cycle temporal resolution required for high-fidelity multidimensional spectroscopy. The CLIMBS configuration geometry delivers attosecond-level delay precision, confirmed by a long-term phase stability of 40 mrad (approximately 16 as at 750 nm), while eliminating beam walk-off, suppressing drift during delay scanning, and supporting a long accessible coherence-time range.

The refractive-index contrast between air and fused silica enables efficient and continuous delay tuning. The use of a thin, isotropic wedge pair within a nearly common-path geometry minimizes the additional dispersion introduced by the pulse shaper compared to birefringent delay lines, while preserving spatial overlap and beam-pointing stability. This architecture also allows independent polarization control of the two pump pulses. This feature will enable polarization schemes in which 2DES spectral features can be enhanced or suppressed so that specific processes may be selectively measured \cite{thyrhaug2018identification}. Together, these features provide high-accuracy, high-speed delay tuning fully compatible with ultrabroadband few-cycle pulses.

Benchmark 2DES measurements on chlorophyll~\textit{a} verify the capabilities of the platform. The MPC excitation spectrum fully covers the $Q_x$ and $Q_y$ absorption bands, ensuring uniform excitation across the entire electronic manifold. The resulting 2D spectra exhibit clean, well-resolved diagonal features at both 200 fs and 800 ps, and the preserved spectral structure across this full temporal window demonstrates that the system maintains phase stability and spectral fidelity from the femtosecond regime to the sub-nanosecond limit.

Overall, the MPC--CLIMBS architecture provides a compact, robust, and experimentally simple route to high-performance multidimensional spectroscopy. By combining ultrabroad bandwidth, attosecond-level delay control, and long-term stability without active stabilization, this platform significantly extends the practical reach of 2DES in both bandwidth and temporal range, and can provide accurate, high-fidelity 2DES measurements across ultrafast coherent and long-lived dynamical regimes.

\bibliography{achemso-demo}

\section*{Associated Content}
\subsection*{Supporting Information}
Supplementary information is available for this paper.

\section*{Author Information}
\subsection*{Corresponding Authors}
*E-mail: bohan.chen@ee.nthu.edu.tw \\
*E-mail: shangda@ee.nthu.edu.tw \\
*E-mail: kai.chen@vuw.ac.nz

\subsection*{Author Contributions}
K.C. conceived the idea and designed the experimental setup. W.-C.F. and B.-H.C. constructed and optimized the system, analyzed the data, and wrote the manuscript. C.-H.L. contributed to project coordination. K.C., H.-S.T., and S.-D.Y. contributed to scientific discussion and provided administrative and general support. All authors discussed the results and commented on the manuscript.

\subsection*{Notes}
The authors declare no competing financial interest.

\section*{Acknowledgments}
The authors thank Prof. I-Chia Chen and members of his group for their assistance with sample preparation.

H.-S.T. acknowledges funding from the Ministry of Education, Singapore (Grant No. T2EP50122-0022).

K.C. acknowledges support from the Victoria University of Wellington FSRG grant (FSRG-ENGRADI-12711) and the Dodd-Walls Centre Investigator Tranche 2.

S.-D.Y. acknowledges funding from the National Science and Technology Council, Taiwan (Grant No. NSTC 112-2112-M-007-018-MY3).

B.-H.C. acknowledges funding from the National Science and Technology Council, Taiwan (Grant No. NSTC 113-2112-M-007-045).

\end{document}